\documentclass[twocolumn,prl,superscriptaddress,preprintnumbers]{revtex4}
\usepackage{graphicx}
\usepackage[figuresright]{rotating}
\usepackage{psfrag}
\newcommand{\be}{\begin{eqnarray}}
\newcommand{\ee}{\end{eqnarray}}
\newcommand{\beq}{\begin{equation}}
\newcommand{\eeq}{\end{equation}}
\newcommand{\ba}{\begin{array}}
\newcommand{\ea}{\end{array}}
\begin{document}
\title{Exclusive annihilation $p\bar p \rightarrow \gamma\gamma$
in a generalized parton picture}
\author{A.~Freund} 
\affiliation{Institut f\"ur Theoretische Physik,
Universit\"at Regensburg, D--93053 Regensburg, Germany}
\author{A.~V.~Radyushkin}
\affiliation{Theory Group, Jefferson Lab, Newport News, VA 23606, USA}
\affiliation{Physics Department, Old Dominion University,
Norfolk, VA 23529, USA}
\author{A.~Sch\"afer} 
\affiliation{Institut f\"ur Theoretische Physik,
Universit\"at Regensburg, D--93053 Regensburg, Germany}
\author{C.~Weiss}
\affiliation{Institut f\"ur Theoretische Physik,
Universit\"at Regensburg, D--93053 Regensburg, Germany}
\begin{abstract}
Exclusive proton--antiproton annihilation into two photons
at large $s$  $(\sim 10 \, {\rm GeV}^2 )$ and $|t|, |u| \sim s$ can be 
described by a generalized parton picture analogous to the 
``soft mechanism'' in wide--angle real Compton scattering. The two photons 
are emitted in the annihilation of a single fast quark and antiquark. 
The matrix element describing the transition of the $p \bar p$ system to a 
$q \bar q$ pair can be related to the timelike proton elastic 
form factors as well as to the quark/antiquark distributions 
measured in inclusive deep--inelastic scattering. 
The reaction could be studied with the proposed $1.5 - 15\, {\rm GeV}$ 
high--luminosity antiproton storage ring (HESR) at GSI.
\end{abstract}
\preprint{JLAB-TH-02-29}
\maketitle
Compton scattering --- both real and virtual --- is one of the main 
sources of information on the structure of the nucleon. In particular,
inclusive deep--inelastic scattering can be viewed as a measurement of the 
imaginary part of the forward virtual Compton amplitude, parametrized 
by the quark and antiquark distributions in the nucleon.
More recently, exclusive processes have been considered, which can be 
described in terms of generalized parton distributions, namely 
deeply--virtual (DVCS) and wide--angle real (WACS) Compton scattering. 
It has been argued that at values of $s \sim \mbox{few} \, {\rm GeV}^2$ 
and $|t|, |u| \sim s$ WACS is dominated by the 
``soft mechanism'' \cite{Radyushkin:1998rt,Diehl:1998kh}.
The Compton scattering occurs off a single quark or antiquark in the 
nucleon. Its emission and absorption by the nucleon is described
by double distributions which can be related to the usual quark/antiquark 
distributions as well as to the elastic form factors of the 
nucleon \cite{Radyushkin:1998rt}.
This approach describes well the existing data \cite{Shupe:vg}, 
including the spin asymmetry of the cross section measured recently 
in the JLAB Hall A experiment \cite{JLAB}. The hard scattering mechanism, 
in which the struck quark rescatters via gluon exchanges of virtuality 
$\sim t$, is relevant only at asymptotically large $s$ and $t$ and cannot 
account for the measured cross section at JLAB 
energies \cite{Radyushkin:1998rt}.
\par
The proposed high--luminosity $1.5-15 \, {\rm GeV}$ antiproton 
storage ring (HESR) at GSI \cite{GSI} would offer an
opportunity to study the Compton process in the crossed channel, 
namely exclusive proton--antiproton annihilation into two photons, 
$p \bar p \rightarrow \gamma\gamma$. In this letter we argue that
this process at large $s$ and $|t|, |u| \sim s$ can also be described
in a generalized parton picture. The two photons are predominantly
emitted in the annihilation of a single ``fast'' quark and antiquark
originating from the proton and antiproton. The new double 
distributions, describing the transition of the $p \bar p$ system to a 
$q \bar q$ pair, can be related to the timelike nucleon form factors; 
by crossing symmetry they are also connected with 
the usual quark/antiquark distributions in the nucleon. 
A similar approach based on light--cone wave functions
has been used for the processes $\gamma\gamma \rightarrow \pi\pi$ 
\cite{Diehl:2001fv} and $\gamma\gamma \rightarrow B\bar B$
\cite{Diehl:2002yh}. Exclusive $p\bar p$
annihilation was also studied in the diquark model \cite{Kroll:1993zx}.
\par
The annihilation process $p(p_1 ) + \bar p (p_2 ) \rightarrow 
\gamma (q_1 ) + \gamma (q_2 )$ is characterized by the invariants
$s = 2 (p_1 p_2) > 0$ (we neglect the proton mass), 
$t = -s \, \sin^2 (\theta/2) < 0$, and $u = - s \,\cos^2 (\theta/2) < 0$,
where $\theta$ is the scattering angle in the center--of--mass frame.
We consider the region where $s$ is much 
larger than typical light hadron masses, $s \sim 10\, {\rm GeV}^2$,
and $|t|, |u| \sim s$, which requires that $\theta$ be sufficiently 
far from $0$ and $\pi$ (wide--angle scattering). 
%
%
\begin{figure}[ht]
\psfrag{mys}{{\Large $s$}}
\psfrag{p_1}{\Large $p_1$}
\psfrag{p_2}{\Large $p_2$}
\psfrag{k_1}{\Large $k_1$}
\psfrag{k_2}{\Large $k_2$}
\psfrag{q_1}{\Large $q_1$}
\psfrag{q_2}{\Large $q_2$}
\psfrag{crossed}{\Large + crossed}
\includegraphics[width=6.7cm,height=3.0cm]{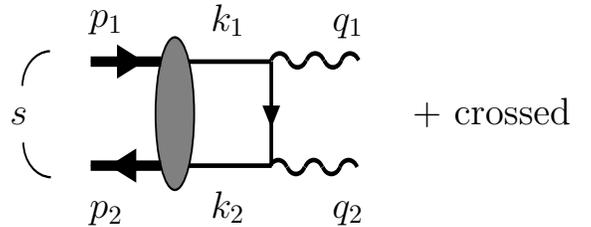}
\caption[]{The ``handbag'' contribution to 
$p \bar p \rightarrow \gamma\gamma$ annihilation.}
\label{fig_handbag}
\end{figure}
\par
In QED, the annihilation process $e^+e^- \rightarrow \gamma \gamma$ 
involves the $t$ (or $u$) channel exchange of a virtual electron/positron
with spacelike four--momentum. In $p \bar p \rightarrow \gamma \gamma $
annihilation in QCD, the exchanged system contains at least three quarks.
At large momentum transfer such an exchange is strongly 
suppressed by the nucleon wave functions. In this situation, the 
most efficient way of accomplishing a large momentum transfer is
the ``handbag'' diagram, Fig.~\ref{fig_handbag}. The amplitude consists
of two parts. In the first part, the proton--antiproton system converts
to a quark--antiquark pair by exchanging a virtual $qq$ 
(``diquark'') system. The nucleon wave functions restrict
the spacelike virtuality of this system to values
of the order of a hadronic scale, $\lambda^2$, related to the size of 
the nucleon. In the second part, the quark--antiquark pair 
annihilates via exchange of a highly virtual quark/antiquark,
much like in QED $e^+ e^- \rightarrow \gamma\gamma$ annihilation.
Neglecting transverse momenta, we expand the momenta of the quark 
and antiquark as
\be
k_1 &=& (1 - x_1 ) p_1 + x_2 p_2 , \\
k_2 &=&      x_1 p_1 + (1 - x_2) p_2 .
\label{active_momenta}
\ee
The variables $x_1$ and $x_2$ obey the ``partonic'' limits $0 < x_{1, 2} < 1$. 
The spacelike virtuality of the exchanged ``diquark'' system is $-x_1 x_2 s$. 
The restriction $|x_1 x_2 s| < \lambda^2$ implies that
for $s \gg \lambda^2$ the value of either $x_1$ or $x_2$ must be 
small, of the order $\lambda^2 / s$. This means that either the 
annihilating quark or the antiquark are 
``fast'', carrying the major part ($\sim 1$) of the proton 
or antiproton momentum. This in turn implies that
the spacelike virtualities of the quark propagators connecting the photon 
vertices in the graphs of 
Fig.~\ref{fig_handbag}, 
$\left[ -x_1 x_2 s + (1 - x_1 - x_2) t \right]$ for the direct and 
$\left[ -x_1 x_2 s + (1 - x_1 - x_2) u \right]$ for the crossed graph, 
are large in the kinematics of wide--angle scattering, $|t|, |u| \sim s$.
Thus, the parton picture of $p \bar p \rightarrow \gamma\gamma$ exclusive 
annihilation emerging from the ``handbag'' graph is self-consistent. 
The inclusion of transverse momenta would change this picture only 
quantitatively.
\par
The so-called hard scattering mechanism would require the rescattering of 
the active partons through gluon exchanges of virtuality $t$. In the case 
of WACS such contributions, which are theoretically dominant in the 
asymptotic region $|t| \rightarrow \infty$, are negligible compared to the 
subasymptotic ``handbag'' contribution at all experimentally relevant values
($t \sim \mbox{few}\; {\rm GeV}^2$), the main reason being that
the hard contributions are numerically suppressed by a factor of 
$(\alpha_s / \pi )^2 \approx 1/100$, see Ref.~\cite{Radyushkin:1998rt}. 
The same can be expected for the annihilation channel.
\par
The crucial ingredient in the calculation of the hadronic amplitude is the 
matrix element describing the amplitude for the conversion of the 
$p\bar p$ system to a $q \bar q$ pair. Following the treatment of 
WACS, we parametrize this matrix element by ``double distributions''
\cite{Radyushkin:1998rt,Radyushkin:1999es,Radyushkin:1999bz},
in which the quark and antiquark momenta are measured in terms of the
proton and antiproton momenta in a symmetric fashion,
using the average and difference, $p = (p_1 + p_2)/2$ and 
$r = p_1 - p_2$:
\be
k_1 &=& (1 + \alpha) p + x r/2 , \\
k_2 &=& (1 - \alpha) p - x r/2 .
\label{active_momenta_alt}
\ee
The new variables $x$ and $\alpha$ are related to the old ones by
$x = 1 - x_1 - x_2, \; \alpha = x_2 - x_1$. We parametrize the 
annihilation--type matrix elements of the quark bilinear 
vector operators as ($a$ denotes the quark flavor)
\be
\lefteqn{\langle 0 | \; \bar\psi_a (-z/2) \gamma_\sigma \psi_a (z/2) \;
| p_1, \lambda_1 ; p_2 \lambda_2 \rangle }
\nonumber \\
&=& \bar v \gamma_\sigma u \;
\int\hspace{-1.5em}\int\limits_{|x| + |\alpha| < 1} \hspace{-1em} 
dx \, d\alpha \; F_a (x, \alpha; s) \; 
e^{-i \alpha (pz) - i x (rz)/2}
\nonumber 
\\
&+& \ldots .
\label{DD_F}
\ee
The matrix element of the axial vector operator ($\gamma_\sigma\gamma_5$) 
is parametrized similarly with a double distribution 
$G_a (x, \alpha; s)$. Here $z$ is a light--like distance, and
$u \equiv u(p_1, \lambda_1 )$ and $\bar v \equiv \bar v(p_2, \lambda_2 )$
are the proton and antiproton spinors. The double distribution
$F_a$ depends on the spectral parameters $x$ and $\alpha$, 
as well as on $s$. Not shown
in Eq.(\ref{DD_F}) for brevity are components of the matrix element
of the type $\epsilon_{\sigma\alpha\beta\gamma} 
\bar v \gamma_\gamma\gamma_5 u$, which play a role in maintaining
electromagnetic gauge invariance of the ``handbag'' amplitude,
similar to the kinematical twist--3 terms in deeply--virtual
Compton scattering \cite{Radyushkin:2000jy}. For simplicity we neglect 
components of the matrix element corresponding to the Pauli and 
pseudoscalar form factors; they should be included in a more complete 
treatment.
\par
The new double distributions are related to the timelike elastic form 
factors of the proton by
\beq
\sum_a e_a
\int\hspace{-1.5em}\int\limits_{|x| + |\alpha| < 1} \hspace{-1em} 
dx \, d\alpha \; F_a (x, \alpha; s) 
\;\; = \;\;  F_1 (s),
\label{reduction_FF}
\eeq
where $F_1 (s)$ is the Dirac vector form factor.
A similar relation holds for the function 
$G_a(x, \alpha ; s)$ and the axial form factor. Furthermore, 
in the limit $s \rightarrow 0$ one can use crossing
symmetry to relate the annihilation--type matrix element 
(\ref{DD_F}) to the corresponding scattering--type
matrix element at $t = 0$, parametrized by the usual double 
distributions of WACS. In particular, this implies
\beq
\int\limits_{-1 + |x|}^{1 - |x|}
\!\!\! d\alpha \; 
F_a (x, \alpha; s = 0) 
\;\; = \;\;  f_a (x ) ,
\label{reduction_PDF}
\eeq
where $f_a (x )$ is the usual unpolarized
parton density in the proton; in conventional notation
$f_a (x) = \theta(x) q_a (x) - 
\theta(-x) \bar q_a (-x)$. In a similar way the
double distribution $G_a (x, \alpha ; s)$ reduces to the
polarized parton density.
\par
To construct an explicit model for the double distributions
we make the ansatz
\beq
F_a (x, \alpha; s) \;\; = \;\; f_a (x) \; h_a (x, \alpha)
\; S_a (x, \alpha; s) .
\label{factorized_ansatz}
\eeq
In the parton density, $f_a (x)$, we take into account only 
the valence quark contribution, {\it i.e.}, we assume it to be of the 
form $f_a (x) = \theta (x) \, q_a (x)$; sea quarks
could readily be included. The factor $h_a(x, \alpha)$
is a profile function, whose integral
over $\alpha$ from $-1 + |x|$ to $1 - |x|$ is
normalized to unity, {\it cf.}\ Eq.(\ref{reduction_PDF}). 
Finally, $S_a (x, \alpha; s)$
is a cutoff function accounting for the $s$--dependence of the double 
distribution, defined such that $S_a (x, \alpha; s = 0) = 1$. 
We model it as (assuming $x > 0$)
\beq
S_a (x, \alpha; s) \;\; = \;\;
\exp \left\{ -\frac{[(1 - x)^2 - \alpha^2] s}
{4 x (1 - x ) \lambda_a^2}
\right\} ,
\label{S_new}
\eeq
or, in terms of the original variables $x_1$ and $x_2$,
\beq
\exp \left[ -\frac{x_1 x_2 s}{(x_1 + x_2)(1 - x_1 - x_2) 
\lambda_a^2} \right] .
\label{S_old}
\eeq
Here $\lambda_a^2$ is a parameter of dimension mass squared.
This cutoff suppresses contributions in which the 
absolute value of the virtuality of the exchanged ``diquark'' system,
$|x_1 x_2 s|$, is large, and also configurations in which 
$x_1 + x_2$ tends to $1$ ($x$ tends to zero). For the double distributions 
parametrizing scattering--type matrix elements in WACS, this cutoff function
can be thought of as the result of an overlap integral of light--cone wave 
functions \cite{Radyushkin:1998rt,Diehl:1998kh}. Such an interpretation is
not possible in the annihilation channel. Here, the cutoff function should
simply be regarded as an effective manner to restrict the virtuality
of the exchanged ``spectator'' system.
\par
For an estimate of the $p\bar p \rightarrow \gamma\gamma$ amplitude
we use the GRV94 LO parametrization of the valence quark densities
at a normalization scale of $1 \, {\rm GeV}^2$ \cite{Gluck:1994uf}. 
We neglect small contributions from strange quarks.
The profile function we take to be of the form
$h_a (x, \alpha) = \delta (\alpha )$; extended profiles have been 
suggested in Refs.\cite{Radyushkin:1999es,Radyushkin:1999bz}. 
For simplicity we choose the cutoff function $S_a (x, \alpha; s)$ 
to be the same for both $u$ and $d$ flavors. The parameter 
$\lambda^2 \equiv \lambda_u^2 = \lambda_d^2$ 
we determine by fitting the integral over the model double distribution
[see Eq.(\ref{reduction_FF})] to the data for the proton vector form factor.
Fitting to the form factor data in the spacelike domain
\cite{Andivahis:1994rq} we obtain $\lambda^2 = 0.7 \, {\rm GeV}^2$. 
For large $s$ the real part of the timelike form factor 
is approximately two times as large as that of the spacelike form factor for 
corresponding large $t$, see Ref.\cite{Bakulev:2000uh} for a discussion. 
In order to account for this effect we multiply the timelike form 
factors entering the double distribution ansatz by a factor of $2$. 
We stress that this should be regarded as a purely phenomenological 
improvement. The axial vector double distribution 
$G_a(x, \alpha; s)$ is modeled analogously, using the GRSV95 LO 
parametrization for the polarized valence quark 
densities \cite{Gluck:1995yr}. For simplicity 
we use here the same cutoff function as in 
the vector double distribution $F_a(x, \alpha; s)$.
\par
With our model for the double distributions we can
compute the $p\bar p \rightarrow \gamma\gamma$ amplitude from the
``handbag'' graphs of Fig.~\ref{fig_handbag}. As the process
is dominated by configurations with 
small $x_1$ and $x_2$ (``fast'' quark and antiquark), we can
simplify the $q \bar q \rightarrow \gamma\gamma$ amplitude by
neglecting contributions of order $x_1 x_2$; keeping them
would result in $\lambda^2 / s$--suppressed contributions
to the $p\bar p \rightarrow \gamma\gamma$ amplitude
which are beyond the accuracy of our approximation. In this way the 
annihilating quark and antiquark are effectively put on mass shell, 
which makes it straightforward to maintain transversality 
of the amplitude (e.m.\ gauge invariance).
The result for the helicity--averaged differential cross section is
\begin{equation}
\frac{d\sigma}{d\cos\theta} 
\;\; = \;\; 
\frac{2\pi \alpha^2_{\rm em}}{s} \; 
\frac{R_V^2 (s) \cos^2 \theta + R_A^2 (s)}{\sin^2 \theta} .
\label{cross}
\end{equation}
The information about the structure of the proton is contained
in generalized form factors
\beq
R_V (s) \;\; \equiv \;\; \sum_a e_a^2
\int\hspace{-1.5em}\int\limits_{|x| + |\alpha| < 1} \hspace{-1em} 
dx \, d\alpha \; 
\frac{F_a (x, \alpha ; s)}{x} ,
\label{R_def}
\eeq
and $R_A (s)$ defined by the corresponding integral over the double
distribution $G_a (x, \alpha ; s)$. For $R_V \equiv R_A \equiv 1$
Eq.(\ref{cross}) would reduce to the Klein--Nishina formula for the
$e^+ e^- \rightarrow \gamma\gamma$ cross section in QED. Note
that our parton picture is applicable only to wide--angle scattering,
so the divergence of the expression in Eq.(\ref{cross}) in the limit 
$\theta \rightarrow 0$ or $\pi$ should be regarded as unphysical.
%
%
\begin{figure}[t]
\psfrag{RAV2}{{\Large $R_{V,A}^2$}}
\psfrag{s/GeV2}{{\Large $s / {\rm GeV}^2$}}
\includegraphics[width=8.4cm,height=5.95cm]{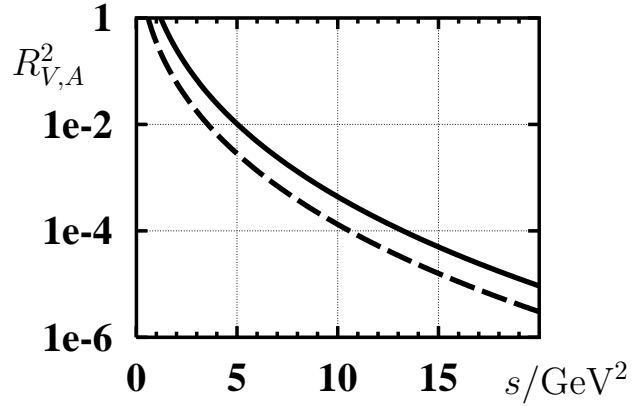}
\caption[]{The squared form factors $R_V^2 (s)$ (solid line)
and $R_A^2 (s)$ (dashed line), as calculated from the double distribution 
model, {\it cf.}\ Eq.(\ref{R_def}).}
\label{fig_sdep}
\end{figure}
\par
Eq.(\ref{cross}) shows that the contribution of the vector operator
to the annihilation cross section, $R_V$, 
is suppressed  at large scattering angle 
relative to that of the axial vector operator, $R_A$, by a factor of
$\cos^2 \theta$. This is different from WACS, where the contribution
from the axial--vector form factor is suppressed \cite{Radyushkin:1998rt}.
\par
Compared to the elastic form factor, 
Eq.(\ref{reduction_FF}), the integrand in Eq.(\ref{R_def}) contains 
an additional factor $1/x$, which is the ``remnant'' of the quark 
propagator in the $q\bar q \rightarrow \gamma\gamma$ amplitude.
Note that the integral nevertheless converges at small $x$, as the 
cutoff function $S_a (x, \alpha; s)$ forces the double distribution
to vanish for $x \rightarrow 0$, 
{\it cf.}\ Eqs.(\ref{factorized_ansatz}) and (\ref{S_new}). More generally, 
the properties of the double distributions ensure that the integrals are
dominated by large values of $x$. The numerical results for the 
squared form factors $R_V (s)$ and $R_A (s)$ are shown in 
Fig.~\ref{fig_sdep}.  
\par
The results for the form factors $R_V (s)$ and $R_A (s)$ turn out to be 
insensitive to the precise value of the normalization scale of the 
parton densities in the double distribution model, 
Eq.(\ref{factorized_ansatz}).
This happens due to the correlation of the $\lambda$ parameter in the 
cutoff function with the normalization scale of the parton 
densities \cite{Vogt:2000ku}. Changing the normalization scale
one must change $\lambda$ such as to refit the form factor data, 
and the two changes compensate each other in the integral 
Eq.(\ref{R_def}). This fact is important for the consistency of 
our approach.
\par
In our simple model the timelike double distributions 
are described by real functions. As a result, the
$p\bar p \rightarrow \gamma\gamma$ amplitude is also real
(the virtuality of the quark propagators in the ``handbag''
graphs of Fig.~\ref{fig_handbag} is always spacelike). 
A more refined treatment should include also
the ``intrinsic'' imaginary part of the double distributions, 
which is related to the imaginary part of the timelike proton
form factor by Eq.(\ref{reduction_FF}).
\par
\begin{figure}[t]
\psfrag{sig/fm2}{{\Large $\sigma / {\rm fm}^2$}}
\psfrag{s/GeV2}{{\Large $s / {\rm GeV}^2$}}
\psfrag{thetarange}{{\Large $45^\circ < \theta < 135^\circ$}}
\includegraphics[width=8.4cm,height=5.95cm]{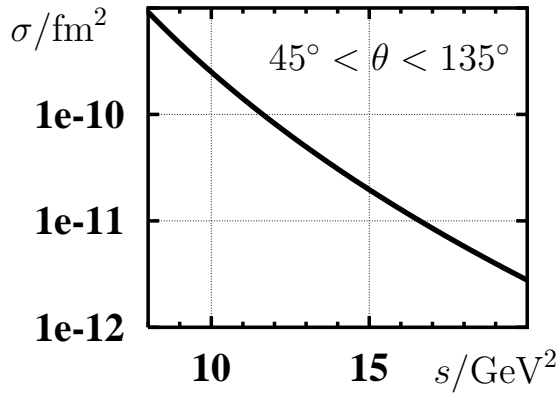}
\caption[]{The $p \bar p \rightarrow \gamma\gamma$ cross section 
integrated over the range $45^\circ < \theta < 135^\circ$,
as a function of $s$.}
\label{fig_sigma}
\end{figure}

It is interesting to estimate the counting rate 
for $p\bar p \rightarrow \gamma\gamma$ annihilation expected for
the proposed $1.5 - 15 \, {\rm GeV}$ antiproton storage ring (HESR) 
at GSI \cite{GSI}. With a solid target the luminosity could be 
as high as $L = 2 \times 10^{32} \, {\rm cm}^{-2} \, {\rm s}^{-1}$. 
Since our parton picture applies only in the kinematical region
where $|t|, |u| \sim s$, we integrate the differential cross section 
(\ref{cross}) over a range 
$\delta < \theta < \pi - \delta$, with $\delta > 0$, excluding small 
scattering angles. For $s = 10 \, {\rm GeV^2}$, a range of 
$45^\circ < \theta < 135^\circ$ corresponds to 
$|t|, |u| > 1.5 \, {\rm GeV^2}$, which seems to be a 
reasonable boundary in view of the experience with wide--angle
Compton scattering \cite{Radyushkin:1998rt}. The cross section 
integrated over this fixed angular range (divided by 2 to account
for the identical particles in the final state) is shown
in Fig.~\ref{fig_sigma} as a function of $s$. At $s = 10 \, {\rm GeV^2}$
our model predicts an integrated cross section of
$0.25 \times 10^{-9}\, {\rm fm}^2$, corresponding to a counting rate of 
$0.5 \times 10^{-3} {\rm sec}^{-1}$, that is $O(10^3)$ events per month. 
The process should thus be measurable with reasonable statistics.
%
%
\par
To summarize, we have outlined a generalized parton picture of
exclusive $p\bar p \rightarrow \gamma\gamma$ annihilation, based
on the ``handbag'' graph and certain assumptions about the soft 
matrix element describing the conversion of the $p\bar p$ system 
to a $q \bar q$ pair. The approximations made in the present treatment 
can be refined in many ways, most notably by including
other components of the soft matrix element (Pauli and 
pseudoscalar form factor), and transverse momenta of the partons. 
Also, the picture proposed here can be applied to
polarization observables, as well as to baryon--antibaryon
production in $\gamma\gamma$ reactions.
\par
We are grateful to M.~D\"uren for starting our interest
in this problem, and to S.~J.~Brodsky, P.~Kroll, K.~Seth, and 
O.~V.~Teryaev for many helpful discussions.
A.~F.\ is supported by an Emmy--Noether Fellowship from 
Deutsche Forschungsgemeinschaft (DFG); 
C.~W.\ by a Heisenberg Fellowship from DFG, A.~V.~R.\ by the
Alexander--von--Humboldt Foundation. This work has been supported 
by DFG, BMBF, and by the US Department of Energy contract
DE-AC05-84ER40150 under which the Southeastern Universities Research 
Association (SURA) operates the Thomas Jefferson Accelerator Facility.
\end{document}